\newcommand{\ket}[1]{\mathop{\left| #1 \right\rangle}\nolimits}

\newcommand{\braket}[2]{\mathop{\left\langle #1 \right|\left. #2\right\rangle}\nolimits}

\newcommand{\enquote}[1]{``#1''}

\documentclass[prl,twocolumn,showpacs,amsmath,amssymb,superscriptaddress]{revtex4-1}
\usepackage{graphicx,bm,color,mathptmx,hyperref} 

\begin{document}
\title{Optimal receiver for binary coherent signals}

\author{Denis Sych}
\affiliation{Max Planck Institute for the Science of Light, G\"{u}nther-Scharowsky-Stra{\ss}e 1/Bau 24, 91058 Erlangen, Germany}
\email{denis.sych@mpl.mpg.de}

\author{Gerd Leuchs}
\affiliation{Max Planck Institute for the Science of Light, G\"{u}nther-Scharowsky-Stra{\ss}e 1/Bau 24, 91058 Erlangen, Germany}
\affiliation{University of Erlangen-Nuremberg, Staudtstra{\ss}e 7/B2, 91058 Erlangen, Germany}
\maketitle

{\bf 
One of the most fascinating aspects of quantum mechanics is the principle impossibility of deterministic errorless discrimination of nonorthogonal signals, such as coherent states. On the one hand, it prevents perfect cloning of quantum states \cite{Dieks:82, Wootters:82} and enables secure communication \cite{Bennett:84}. On the other hand, it makes a grand challenge to reach the ultimate measurement precision. Although the minimum possible error rate (the Helstrom bound \cite{Helstrom:67}) has been known for almost five decades, there is no practical way to achieve it \cite{Kennedy:73,Dolinar:73,Sasaki:96,Cook:07,Chen:12,Becerra:13}. Developing the realistic optimal measurement strategies to attain the Helstrom bound is of utmost importance for high-precision applications, long-distance free-space and optical fiber communication, gravitational wave detection, optical sensing in biology and medicine, to name a few. In this work, we show an optimal receiver for coherent states which admits a relatively simple technological implementation. The receiver is based on multichannel splitting of the signal, followed by feed-forward signal displacement and photon detection.}

Of all pure quantum states, coherent states are most robust against loss hence their importance for a wide range of applications and the need for the best possible detection strategies. In the simplest scenario, one has to discriminate between two coherent states $\ket{\alpha}$ and $\ket{-\alpha}$ that have equal {\em a priori} probabilities, i.e. to identify the binary signal \cite{Helstrom:76,Bergou:10}. Due to the nonzero overlap $|\braket{\alpha}{-\alpha}|^2=e^{-4\alpha^2}$, there is a certain error rate, which depends on the measurement strategy. In the middle 60's, Helstrom found the minimum possible error rate
$\epsilon_{\rm Helstrom}=\frac{1}{2}\left(1-\sqrt{1-e^{-4\alpha^2}}\right)$, but provided no specific recipe how to construct a practical setup to achieve it  \cite{Helstrom:67}. 

\begin{figure}[h]
\includegraphics[width=0.95\columnwidth]{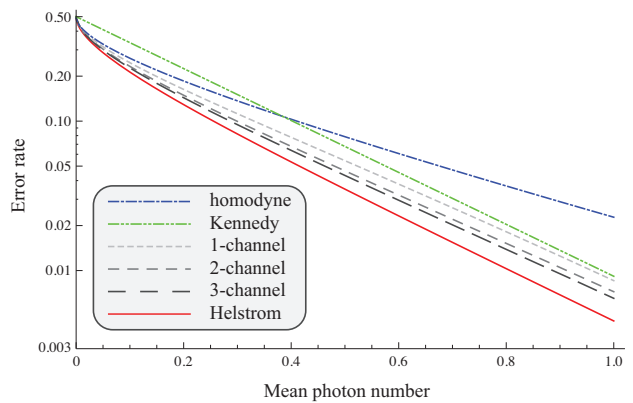}
\caption{\label{fig:e2} {\bf Receiver performance.} Error rate $\epsilon$ (logarithmic scale) as a function of mean photon number $m$ (linear scale) for various receivers: homodyne (blue); Kennedy (green); 1-, 2-, and 3-channel optimal displacement (light gray, gray, and dark gray); and the Helstrom bound (red)}
\end{figure}

Conventional coherent optical receivers based on homodyne detection operate at the shot-noise limit, which leaves an exponential gap in the error rate well above the Helstrom bound (see Fig.~\ref{fig:e2} for comparative performance of various receivers). To outperform the homodyne receiver, Kennedy proposed a receiver which relies on state displacement and single-photon detection \cite{Kennedy:73}. For sufficiently strong signals, the Kennedy receiver outperforms homodyne detection, but it doesn't reach the Helstrom bound. 

In quest of optimal detection, Dolinar improved the Kennedy receiver by adding adaptive feed-back control and time-dependent displacement \cite{Dolinar:73}. Theoretically, the Dolinar receiver reaches the Helstrom bound, but its practical realization is restrained due to several reasons. First, in order to achieve a sufficiently large number of adaptive iterations, the bandwidth of the detector and electronic components must be much larger than the symbol repetition rate. Also the feed-back delay, limited by the physical size of the control circuit, must be much shorter than the symbol window. These aspects restrict the practical applicability of the Dolinar receiver for the telecom use, since the low error rate is inevitably connected with the low symbol repetition rate. In practice, it makes more sense to use other detection strategies which posses a higher error rate but at the same time allow a higher repetition rate. Second, continuous measurement of the signal and fast feed-back control impose strict requirements on the performance of the detector, namely, high quantum efficiency, low dark count rate, short dead time, and small timing jitter, which is very hard to realize in an experiment \cite{Achilles:03,Achilles:04,Jiang:07,Kardynal:08,Eisaman:11,Fukuda:11,Akiba:12,Calkins:13,Marsili:13}.

In this work, we demonstrate a quantum receiver scheme which combines the best features of the Kennedy and Dolinar receivers: a simple experimental realization of the former, and the optimal performance of the latter. The receiver allows to approach the Helstrom bound arbitrarily closely, but avoids the above mentioned problems of the Dolinar receiver. In particular, impracticable oversampling in time and instantaneous feed-back are replaced with much more feasible spatial diversity and delayed feed-forward.

\begin{figure}[h]
\includegraphics[width=0.95\columnwidth]{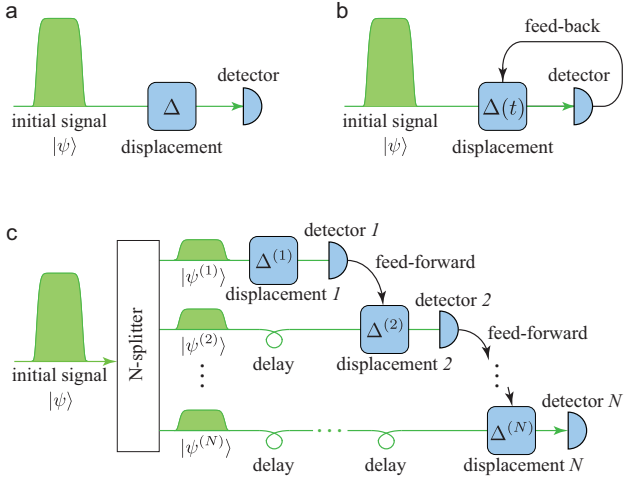}
\caption{\label{fig:receivers} {\bf Detection of binary coherent signals.} {\bf a}, The Kennedy receiver displaces the initial signal and measures it with a single-photon detector. {\bf b}, The Dolinar receiver performs a real-time feed-back control of displacement $\Delta(t)$ to null the initial signal. {\bf c}, The feed-forward receiver splits the initial signal into several channels. Quantum state in each channel is displaced by $\Delta^{(k)}$ and then measured with a photon-number-resolving detector. Via feed-forward control, the value of displacement $\Delta^{(k)}$ in each channel depends on the measurement results in the preceding channels. Prior to the displacement, the signal is delayed to match the feed-forward speed.}
\end{figure}

To get an insight into the optimal discrimination strategy, let us start with the Kennedy receiver (see Fig.~\ref{fig:receivers}a). The displacement operation is performed by mixing the signal with a reference field on an almost transparent beamsplitter. The reference field is chosen such that after the beamsplitter the initial signal $\ket{\psi}=\{\ket{\alpha},\ket{-\alpha}\}$ is transformed to $\ket{\psi}=\{\ket{\alpha+\Delta},\ket{-\alpha+\Delta}\}$. The value of displacement is equal to the signal amplitude $\Delta=\alpha$, thus the displaced states are $\{\ket{2\alpha},\ket{0}\}$. In the case of a detector click, the receiver unambiguously discriminates the first state and selects $\ket{\alpha}$. Otherwise, the receiver selects the state $\ket{-\alpha}$, which can lead to the discrimination error.

Dolinar developed the idea further. He proposed to adjust the amplitude and phase of the reference displacement field such that after the beamsplitter the initial signal $\ket{\psi(t)}$ is dynamically transformed to $\ket{\psi(t)+\Delta(t)}$ (see Fig.~\ref{fig:receivers}b). The absolute value of displacement $\Delta(t)$ at time $t$ is given by a specially tailored function $f(t)$, and the choice of the sign depends on the most probable detection hypothesis at time $t$: $\Delta(t)=f(t)$ corresponds to $\ket{-\alpha}$, and $\Delta(t)=-f(t)$ corresponds to $\ket{\alpha}$. At the beginning of the signal, the initial detection hypothesis is chosen according to the largest prior probability, or chosen randomly in case of equal priors. Every time the detector clicks, the feed-back control switches between the detection hypotheses, and flips the sign of $\Delta(t)$. The final decision on the signal state is determined by the last detection hypothesis at the end of the signal. As we pointed out, the main difficulty to realize the Dolinar receiver in an experiment is impracticable instantaneous feed-back control.

The schematic setup of the optimal receiver proposed in this work is shown in Fig.~\ref{fig:receivers}c. The initial binary signal $\ket{\psi}=\{\ket{\alpha_1},\ket{\alpha_2}\}$ is split into $N$ channels $(\ket\psi=\ket{\psi^{(1)}}\otimes\ket{\psi^{(2)}}\otimes\ldots\otimes\ket{\psi^{(N)}})$. Quantum state $\ket{\psi^{(k)}}$ in each channel is displaced to $\ket{\psi^{(k)}+\Delta^{(k)}}$ and then measured by a photon counting detector. The final decision on the signal state is made after measuring the last channel $N$. The choice of parameters (such as splitting ratio, number of channels, type of detectors, value of displacement, and data processing algorithm) determines several possible signal measurement strategies. In the simplest scenario, the signal splitting is homogenous ($\ket{\psi^{(1)}}=\ket{\psi^{(2)}}=\ldots=\ket{\psi^{(N)}}=\ket{{\rm \psi}/\sqrt{N}}$), all displacements $\Delta^{(k)}$ have equal amplitudes and variable phases. In a more advanced approach, the splitting is inhomogenous, the phase and the amplitude of displacements $\Delta^{(k)}$ may dynamically depend on the measurement results in the previous channels.

Let us start with the single-channel case ($N=1$). We assume phase-shift modulation of the signal states ($\ket{\alpha_{1}}=\ket{\alpha}$ and $\ket{\alpha_{2}}=\ket{-\alpha}$) which have prior probabilities $p_1$ and $p_2=1-p_1$, respectively. In fact, for the purpose of state discrimination, the exact form of binary encoding is not important, because the only relevant parameter is the overlap $|\braket{\alpha_1}{\alpha_2}|^2=e^{-|\alpha_1-\alpha_2|^2}$. Phase-shift modulation $\ket{\alpha_{1,2}}=\ket{\pm\alpha}$ is equivalent to on-off modulation $(\ket{\alpha_{1}}=\ket{0}$, $\ket{\alpha_{2}}=\ket{2\alpha})$ up to the displacement $\Delta=\alpha$. Along with the amplitude $\alpha$ we will also use the mean photon number $m=\alpha^2$.

If $p_1\leq p_2$, we perform a displacement $\Delta=\alpha+\beta$, which has an increment $\beta$ over the ``exact nulling'' displacement $\Delta=\alpha$. In the opposite case ($p_1>p_2$) we apply the same displacement with the negative sign $(\Delta=-\alpha-\beta)$. 
The displaced states are then
\begin{equation}
\begin{array}{ll}
\ket{\alpha_1}=\ket{2\alpha+\beta},  \ket{\alpha_2}=\ket{\beta}, & \quad {\rm when}\quad  p_1\leq p_2\\
\ket{\alpha_1}=\ket{-\beta},  \ket{\alpha_2}=\ket{-2\alpha-\beta}, & \quad {\rm when}\quad  p_1>p_2.
\end{array}
\label{eq:displ}
\end{equation}

A photon-number-resolving detector gives us a discrete set of outcomes (number of photons) $n=0,1,2,\ldots$, that corresponds to projection of the signal state $\ket\psi$ onto Fock basis $\{\ket n\}$. For a coherent state $\ket\alpha$, the probability $P(n,\alpha)$ of an outcome $n$ is given by the Poisson distribution $P(n,\alpha)=|\alpha|^{2n}e^{-|\alpha|^2}/n!$. 

Our discrimination strategy relies on the maximum posterior probability: for an outcome $n$, we chose the signal state $\ket{\alpha_i}$ that has the maximum value $p_i P(n,\alpha_i)$. Provided the outcome $n$, the correct identification of the signal has probability $f_n=p_i P(n,\alpha_i)/(p_1 P(n,\alpha_1)+p_2 P(n,\alpha_2))$. Taking into account that probability $P_n$ to obtain the outcome $n$ is $P_n=p_1 P(n,\alpha_1)+p_2 P(n,\alpha_2)$, we get the average error rate 
\begin{equation}
\epsilon=1-\sum\limits_n \max\limits_{i}[p_i P(n,\alpha_i)].
\label{eq:err1}
\end{equation} 

Clearly, the displacement operation (\ref{eq:displ}) doesn't change the overlap $|\braket{\alpha_1}{\alpha_2}|^2$, but it does change the probability distribution $P(n,\alpha_i)$, therefore, it affects the error rate (\ref{eq:err1}). As we show below, the optimal displacement increment $\beta_{opt}$ must be non-negative to minimize the error rate. The first near-optimal receiver proposed by Kennedy implies $\beta=0$ \cite{Kennedy:73}, and several recent modifications of the Kennedy receiver use $\beta>0$ \cite{Takeoka:08,Wittmann:08,Wittmann:10prl,Wittmann:10pra,Wittmann:10jmo,Tsujino:10,Tsujino:11}.

\begin{figure}[h]
\includegraphics[width=0.95\columnwidth]{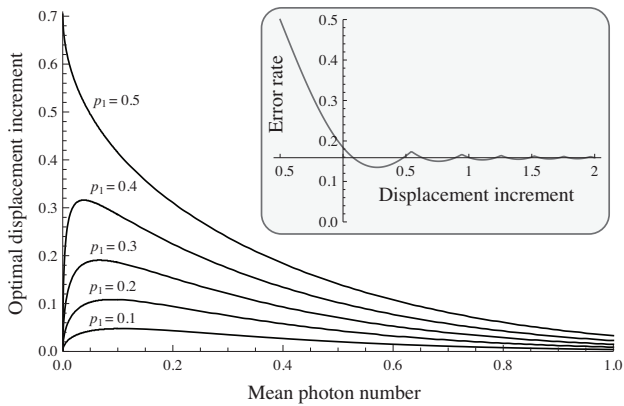}
\caption{\label{fig:bopt} {\bf Single-channel receiver optimization.} Optimal displacement increment $\beta_{opt}$ as a function of mean photon number $m$ for various prior signal probabilities $p_1=$ 0.1, 0.2, 0.3, 0.4, and 0.5. {\em Inset figure}: The error rate $\epsilon$ as a function of displacement increment $\beta$ for the unbiased signal ($p_1=p_2=1/2$) with the mean photon number $m=0.25$. Optimal displacement increment $\beta_{opt}$ corresponds to the first local minimum of $\epsilon$, and the limit of large $\beta$ shows the homodyne detection error rate $\epsilon_{\rm Homodyne}\simeq 0.159$.}
\end{figure}

Without loss of generality, we assume $p_1\leq p_2$. The error rate $\epsilon$ as a function of $\beta$ is shown in the inset to Fig.~\ref{fig:bopt}. Surprisingly, this is not a monotonic function. For a given signal amplitude $\alpha$ and a displacement increment $\beta$, there is a certain discrimination threshold $n^*$ such that all measurement outcomes $n\geq n^*$ are assigned to the state $\alpha_1$, whilst the rest outcomes $n<n^*$ to the state  $\alpha_2$. Infinitely many local minima of $\epsilon(\beta)$ correspond to different values of $n^*$. Global minimization of the error rate (\ref{eq:err1}) with respect to the displacement can be done by taking the first root of equation $\frac{{\rm d}\epsilon}{{\rm d}\beta}=0$, which corresponds to the discrimination threshold $n^*=1$. It means, that we can relax the condition of photon-number-resolving detection to ``on-off'' detection. The discrimination strategy can be simplified: if the detector registers at least one photon (``on''), we chose the state $\ket{\alpha_1}$, otherwise (``off'') we chose the state $\ket{\alpha_2}$. For this ``on-off'' strategy, the error rate is 
\begin{equation}
\epsilon_{\rm on-off}=p_1 e^{-|2\alpha+\beta|^2}+p_2 (1-e^{-|\beta|^2}).
\label{eq:onoff}
\end{equation}

The results of optimization of the displacement increment $\beta$ are shown in Fig.~\ref{fig:bopt}. For discrimination of strong signals ($m\gg1$), optimal displacement increment $\beta_{opt}$ tends to zero. This explains why the Kennedy receiver ($\beta=0$), which provides the error rate $\epsilon_{\rm Kennedy}=p_1e^{-4m}$, demonstrates relatively good performance in the domain $m\gg1$. The opposite extreme case~--- discrimination of very weak signals ($m\ll1$)~--- has rather counterintuitive optimization: $\beta_{opt}$ also tends to zero, except of the case of the unbiased signal ($p_1=p_2=1/2$), where it tends to $1/\sqrt{2}\simeq 0.7$. 

For comparison, the homodyne receiver can be regarded as the ultimate case of very large displacement $\beta\gg1$ and photon-number-resolving detection, which leads to the error rate $\epsilon_{\rm homodyne}=(1-{\rm erf}(\sqrt{2}\alpha))/2$, where ${\rm erf}(x)=2\int_0^x e^{-t^2}{\rm d}t/\sqrt \pi$ is the error function.

Now we proceed with the generic multichannel case. As we can see from the results of single-channel optimization, the optimal displacement non-trivially depends on the prior probabilities and amplitudes of the signal states. If we split first the initial signal $\ket\psi$ to $N$ channels, we can perform then an optimal state displacement in each channel, conditioned on the particular measurement results in the previous channels. More specifically, we exploit feed-forward Bayessian update of signal probability distribution, i.e. take posterior probabilities of the signal states in each channel $k$ as prior probabilities for the subsequent channel $k+1$, and perform an optimized displacement $\Delta^{(k)}$ to match them \cite{Becerra:11,Izumi:12}. Prior to the displacement, the signal has a proper temporal delay between the channels in order to to match the realistic  speed of the detectors and feed-forward electronics.

Similarly to the single-channel case (\ref{eq:displ}), we apply positive displacement $\Delta^{(k)}=\alpha^{(k)}+\beta^{(k)}$ when $p_1^{(k)}\leq p_2^{(k)}$, and negative displacement $\Delta^{(k)}=-\alpha^{(k)}-\beta^{(k)}$ otherwise. Provided a particular measurement result $n^{(k)}$ in the channel $k$, the conditional transformation of probabilities $p^{(k)}_i$ is
$p^{(k+1)}_i=p^{(k)}_i P(n^{(k)},\alpha^{(k)}_i)/\left(p^{(k)}_1 P(n^{(k)},\alpha^{(k)}_1)+p^{(k)}_2 P(n^{(k)},\alpha^{(k)}_2)\right)$,
where the measurement outcomes $n^{(k)}$ have the Poisson probability distribution $P(n^{(k)},\alpha^{(k)}_i)$.

The feed-forward measurement procedure starts in the first channel with the initial probability distribution $p^{(1)}_i=\{p_1^{(1)},p_2^{(1)}\}=\{p_1,p_2\}$. After the first displacement $\Delta^{(1)}$ and the first measurement, we get a result $n^{(1)}$, which defines prior probability distribution $p^{(2)}_i=\{p_1^{(2)},p_2^{(2)}\}$ for the second channel. After the second displacement $\Delta^{(2)}$ and the second measurement, we get a result $n^{(2)}$, which leads to $p^{(3)}_i=\{p_1^{(3)},p_2^{(3)}\}$, etc. In the last channel $N$, we get the final measurement result $n^{(N)}$ that determines the output of the receiver according to the largest term of the final probability distribution $p^{(N+1)}_i=\{p^{(N+1)}_1,p^{(N+1)}_2\}$.

Important to note that joint optimization of all channels differs from individual channel optimization and involves photon-number-resolving detection and inhomogeneous signal splitting. In general, the larger the number of channels, the smaller the error rate. For any number of channels, we find the optimal discrimination threshold $n^*=1$, similarly to the single-channel case. In the asymptotic case $N\rightarrow\infty$, the error rate tends to the Helstrom bound, and optimization can be done in a simplified scenario, which involves only homogeneous channel splitting ($\ket{\psi^{(k)}}=\ket{\psi/\sqrt{N}}$) and ``on-off'' detection (i.e. only two types of detection events: $n=0$ (``off'') and $n>0$ (``on'')). Fig.~\ref{fig:e2} shows the results of optimization. 

For a large number of channels, a convenient way to enumerate the channels is to use a normalized index $\kappa=k/N$. In the most interesting case of equiprobable states ($p_1=p_2=1/2$), the optimal displacement increment $\beta^{(\kappa)}$ takes the form analogous to the Dolinar receiver
\begin{equation}
\beta^{(\kappa)}=\frac{1-\sqrt{1-e^{-4m\kappa}}}{\sqrt{1-e^{-4m\kappa}}}\sqrt{m}.
\label{eq:infch}
\end{equation}

The optimal displacement control strategy for a given channel $\kappa$ turns out to be very simple: the displacement increment is given by Eq.~\ref{eq:infch}, and the most likely state hypothesis depends on the parity of the total number of ``on'' events in the previous channels. For even number of ``on'' events, the most likely state is $\ket{-\alpha}$, thus we apply a ``positive'' displacement $\Delta^\kappa=\alpha^{(\kappa)}+\beta^{(\kappa)}$, otherwise we choose state $\ket{\alpha}$ and apply a ``negative'' displacement $\Delta^{(\kappa)}=-\alpha^{(\kappa)}-\beta^{(\kappa)}$. The final discrimination decision is given by parity of the overall number of ``on'' events in all channels: $\ket{\alpha}$ for odd number and $\ket{-\alpha}$ for even number.

We point out that the proposed receiver has numerous major advantages compared to the Dolinar receiver. First, it can work at the repetition rate corresponding to one measurement cycle of a single ``on-off'' detector, and it doesn't depend on the number of channels. Indeed, the total measurement time is proportional to the number of channels, but after one measurement in the first channel, the signal is processed only in the subsequent channels. Thus, the first detector can proceed with the next signal, while the preceding signals have still being processed in the other channels. Provided the same technical components, this receiver can work at much higher repetition rate than the Dolinar receiver, which leads to higher communication rate. Second, the receiver doesn't require instantaneous feed-forward control, i.e. the electronic control bandwidth can be reduced down to the signal repetition rate. A temporal delay between the channels can be simply realized with a sufficiently long optical fiber. For an electronic bandwidth of 1 GHz, the sufficient delay of 1 nanosecond requires only 20 centimeters of standard optical fiber, which causes almost no losses of the signal ($4*10^{-5}$ dB or $0.001\%$ for the telecom fiber with 0.2 dB/km losses), apart from insertion loss. Third, detectors can be operated in the gated regime, which provides higher stability to the dark counts and dead time. Moreover, the receiver has better stability to overall operational inaccuracy. The Dolinar receiver changes the displacement signal completely after every click, accumulating all kinds of inaccuracies and inperfections. In our case, the signal is displaced in each channel separately and only once. Fourth, there is no need for fast temporal shaping of the displacement control signal, since temporal profiles of the control pulse and the signal pulse must be the same to guarantee interference.

In conclusion, we proposed a quantum receiver scheme which is based on multichannel signal splitting followed by state displacement and photon counting measurement. The scheme has several free parameters (number of channels, splitting ratio, value of displacement,  type of detectors, and feed-forward control algorithm) that allow flexible optimization and make the receiver both versatile and practical. We have shown that optimization of the receiver for minimum error discrimination of binary coherent states leads to the Helstrom bound. We anticipate our theoretical findings to motivate future work towards their experimental implementations in various applications as well as theoretical development of optimal receiver schemes aimed at other types of signals and discrimination strategies.

{\bf Acknowledgements} The authors thank Georgy Onishchukov and Luis L. S\'{a}nchez-Soto for useful discussions and comments on the manuscript.


\begin{thebibliography}{10}

\bibitem{Dieks:82}
D.~Dieks, \enquote{Communication by EPR devices,} Phys. Lett. A \textbf{92},
  271 (1982).

\bibitem{Wootters:82}
W.~K. Wootters and W.~H. Zurek, \enquote{A single quantum cannot be cloned,}
  Nature \textbf{299}, 802 (1982).

\bibitem{Bennett:84}
C.~H. Bennett and G.~Brassard, \enquote{Quantum cryptography: Public key
  distribution and coin tossing,} in \enquote{Proc. IEEE Int. Conf. on
  Computers, Systems ad Signal Processing,}  (IEEE, New York, 1984), p. 175.

\bibitem{Helstrom:67}
C.~W. Helstrom, \enquote{Detection theory and quantum mechanics,} Inf. and
  Control \textbf{10}, 254--291 (1967).

\bibitem{Kennedy:73}
R.~Kennedy, \enquote{A near-optimum receiver for the binary coherent state
  quantum channel,} MIT Res. Lab. Electron. Quart. Progr. Rep. \textbf{108},
  219--225 (1973).

\bibitem{Dolinar:73}
S.~Dolinar, \enquote{An optimum receiver for the binary coherent state quantum
  channel,} MIT Res. Lab. Electron. Quart. Progr. Rep. \textbf{111}, 115--120
  (1973).

\bibitem{Sasaki:96}
M.~Sasaki and O.~Hirota, \enquote{Quantum decision scheme with a unitary
  control process for binary quantum-state signals,} Phys. Rev. A \textbf{54},
  2728--2736 (1996).

\bibitem{Cook:07}
R.~L. Cook, P.~J. Martin, and J.~M. Geremia, \enquote{Optical coherent state
  discrimination using a closed-loop quantum measurement,} Nature \textbf{446},
  774--777 (2007).

\bibitem{Chen:12}
J.~Chen, J.~L. Habif, Z.~Dutton, R.~Lazarus, and S.~Guha, \enquote{Optical
  codeword demodulation with error rates below the standard quantum limit using
  a conditional nulling receiver,} Nature Phot. \textbf{6}, 374--379 (2012).

\bibitem{Becerra:13}
F.~E. Becerra, J.~Fan, G.~Baumgartner, J.~Goldhar, J.~T. Kosloski, and
  A.~Migdall, \enquote{Experimental demonstration of a receiver beating the
  standard quantum limit for multiple nonorthogonal state discrimination,}
  Nature Phot. \textbf{7}, 147--152 (2013).

\bibitem{Helstrom:76}
C.~W. Helstrom, \emph{Quantum Detection and Estimation Theory} (Academic Press,
  New York, 1976).

\bibitem{Bergou:10}
J.~A. Bergou, \enquote{Discrimination of quantum states,} J. Mod. Opt.
  \textbf{57}, 160--180 (2010).

\bibitem{Achilles:03}
D.~Achilles, C.~Silberhorn, C.~{\'S}liwa, K.~Banaszek, and I.~A. Walmsley,
  \enquote{Fiber-assisted detection with photon number resolution,} Opt.
  Letters \textbf{28}, 2387--2389 (2003).

\bibitem{Achilles:04}
D.~Achilles, C.~Silberhorn, C.~{\'S}liwa, K.~Banaszek, I.~A. Walmsley, M.~J.
  Fitch, B.~C. Jacobs, T.~B. Pittman, and J.~D. Franson,
  \enquote{Photon-number-resolving detection using time-multiplexing,} J. Mod.
  Opt. \textbf{51}, 1499--1515 (2004).

\bibitem{Jiang:07}
L.~A. Jiang, E.~A. Dauler, and J.~T. Chang, \enquote{Photon-number-resolving
  detector with 10bits of resolution,} Phys. Rev. A \textbf{75}, 062325 (2007).

\bibitem{Kardynal:08}
B.~E. Kardyna\l, Z.~L. Yuan, and A.~J. Shields, \enquote{An
  avalanche-photodiode-based photon-number-resolving detector,} Nature Phot.
  \textbf{2}, 425--428 (2008).

\bibitem{Eisaman:11}
M.~D. Eisaman, J.~Fan, A.~Migdall, and S.~V. Polyakov, \enquote{Invited review
  article: Single-photon sources and detectors,} Rev. Sci. Instrum.
  \textbf{82}, 071101 (2011).

\bibitem{Fukuda:11}
D.~Fukuda, G.~Fujii, T.~Numata, K.~Amemiya, A.~Yoshizawa, H.~Tsuchida,
  H.~Fujino, H.~Ishii, T.~Itatani, S.~Inoue, and T.~Zama,
  \enquote{Titanium-based transition-edge photon number resolving detector with
  98\% detection efficiency with index-matched small-gap fiber coupling,} Opt.
  Express \textbf{19}, 870--875 (2011).

\bibitem{Akiba:12}
M.~Akiba, K.~Inagaki, and K.~Tsujino, \enquote{Photon number resolving sipm
  detector with 1 GHz count rate,} Opt. Express \textbf{20}, 2779--2788 (2012).

\bibitem{Calkins:13}
B.~Calkins, P.~L. Mennea, A.~E. Lita, B.~J. Metcalf, W.~S. Kolthammer,
  A.~Lamas-Linares, J.~B. Spring, P.~C. Humphreys, R.~P. Mirin, J.~C. Gates,
  P.~G.~R. Smith, I.~A. Walmsley, T.~Gerrits, and S.~W. Nam, \enquote{High
  quantum-efficiency photon-number-resolving detector for photonic on-chip
  information processing,} Opt. Express \textbf{21}, 22657--22670 (2013).

\bibitem{Marsili:13}
F.~Marsili, V.~B. Verma, J.~A. Stern, S.~Harrington, A.~E. Lita, T.~Gerrits,
  I.~Vayshenker, B.~Baek, M.~D. Shaw, R.~P. Mirin, and S.~W. Nam,
  \enquote{Detecting single infrared photons with 93\% system efficiency,}
  Nature Phot. \textbf{7}, 210--214 (2013).

\bibitem{Takeoka:08}
M.~Takeoka and M.~Sasaki, \enquote{Discrimination of the binary coherent
  signal: Gaussian-operation limit and simple non-gaussian near-optimal
  receivers,} Phys. Rev. A \textbf{78}, 022320 (2008).

\bibitem{Wittmann:08}
C.~Wittmann, M.~Takeoka, K.~N. Cassemiro, M.~Sasaki, G.~Leuchs, and U.~L.
  Andersen, \enquote{Demonstration of near-optimal discrimination of optical
  coherent states,} Phys. Rev. Lett. \textbf{101} (2008).

\bibitem{Wittmann:10prl}
C.~Wittmann, U.~L. Andersen, M.~Takeoka, D.~Sych, and G.~Leuchs,
  \enquote{Demonstration of coherent-state discrimination using a
  displacement-controlled photon-number-resolving detector,} Phys. Rev. Lett.
  \textbf{104}, 100505 (2010).

\bibitem{Wittmann:10pra}
C.~Wittmann, U.~L. Andersen, M.~Takeoka, D.~Sych, and G.~Leuchs,
  \enquote{Discrimination of binary coherent states using a homodyne detector
  and a photon number resolving detector,} Phys. Rev. A \textbf{81}, 062338
  (2010).

\bibitem{Wittmann:10jmo}
C.~Wittmann, U.~L. Andersen, and G.~Leuchs, \enquote{Discrimination of optical
  coherent states using a photon number resolving detector,} J. Mod. Opt.
  \textbf{57}, 213--217 (2010).

\bibitem{Tsujino:10}
K.~Tsujino, D.~Fukuda, G.~Fujii, S.~Inoue, M.~Fujiwara, M.~Takeoka, and
  M.~Sasaki, \enquote{Sub-shot-noise-limit discrimination of on-off keyed
  coherent signals via a quantum receiver with a superconducting transition
  edge sensor,} Opt. Express \textbf{18}, 8107--8114 (2010).

\bibitem{Tsujino:11}
K.~Tsujino, D.~Fukuda, G.~Fujii, S.~Inoue, M.~Fujiwara, M.~Takeoka, and
  M.~Sasaki, \enquote{Quantum receiver beyond the standard quantum limit of
  coherent optical communication,} Phys. Rev. Lett. \textbf{106}, 250503
  (2011).

\bibitem{Becerra:11}
F.~E. Becerra, J.~Fan, G.~Baumgartner, S.~V. Polyakov, J.~Goldhar, J.~T.
  Kosloski, and A.~Migdall, \enquote{M-ary-state phase-shift-keying
  discrimination below the homodyne limit,} Phys. Rev. A \textbf{84} (2011).

\bibitem{Izumi:12}
S.~Izumi, M.~Takeoka, M.~Fujiwara, N.~D. Pozza, A.~Assalini, K.~Ema, and
  M.~Sasaki, \enquote{Displacement receiver for phase-shift-keyed coherent
  states,} Phys. Rev. A \textbf{86}, 042328 (2012).

\end{thebibliography}
\end{document}